\documentstyle[12pt]{article}
\newcommand{\be}{\begin{equation}}
\newcommand{\ee}{\end{equation}}
\newcommand{\bea}{\begin{eqnarray}}
\newcommand{\eea}{\end{eqnarray}}
\newcommand{\sptwo}{1.4}
\newcommand{\doublespace}{\edef\baselinestretch{\sptwo}\Large\normalsize}
\begin{document}
\hspace*{\fill} PURD-TH-95-04\\

\vspace{0.5in}
\begin{center}
{\large\bf AN OPERATOR ANALYSIS OF A SUPERSYMMETRIC EFFECTIVE THEORY\\}
\end{center}
{}~\\
\begin{center}
{\bf M.A. Walker}\footnote{Supported in part by a Purdue Research Foundation
Research Grant.}\\
{\it Department of Physics}\\
{\it Purdue University}\\
{\it West Lafayette, IN 47907-1396}
{}~\\
{}~\\
\end{center}
{}~\\
{}~\\
\begin{center}
{\bf Abstract}
\end{center}
An analysis of a $SU(2)_L \times SU(2)_R$ invariant, supersymmetric effective
theory is given. The resulting leading and next to leading independent
invariants are stated in terms of the underlying Killing vectors and K\"ahler
potential. The appendices are devoted to the relationship between this
geometrical point of view and the standard unitary matrix formulation.
\pagebreak
\doublespace

Effective Lagrangian techniques have been increasingly employed to describe
lower energy behavior of models assuming there are strongly interacting
dynamics underlying the theory. In this way, deviations from low energy
behavior can be parametrized. A prominent example of this occurs in the
electroweak chiral Lagrangian. In this case, if the electroweak symmetry
breaking occurs because of a heavy Higgs bosonic sector, a low energy expansion
can be made in powers of momentum.

This type of expansion has been studied in detail with invariants calculated to
$O(p^4)$$^{[\ref{long},\ref{appel}]}$. In this case, since the model respects a
$SU(2)_L \times U(1)$ symmetry, the fields can be grouped into a dimensionless
two by two unitary matrix and its complex conjugate. These matrices have
properties that allow a reasonably simple determination of the invariants and
indeed, two $O(p^2)$ and fourteen $O(p^4)$ terms have been enumerated for the
gauged symmetry. For the case of a global $SU(2)_L \times SU(2)_R$ symmetry,
the number of invariants drops to one $O(p^2)$ and two $O(p^4)$ terms.

However in supersymmetric models, the fields become complex chiral superfields.
This presents a complication in the scheme presented by
Longhitano$^{[\ref{long}]}$ to categorize independent invariants since the
convenient matrix properties exhibited in the electroweak case no longer hold.
For the case of $SU(2)_L \times SU(2)_R$ global symmetry broken to the vector
subgroup $SU(2)_V$, one can proceed analogous to Longhitano by defining the
unitary matrices
\bea
U(\vec{x},\theta,\bar{\theta}) &=& fexp(i \vec{T}\cdot \vec{\xi})\nonumber\\
U^{-1}(\vec{x},\theta,\bar{\theta}) &=& fexp(-i \vec{T} \cdot\vec{\xi})\,
\eea
where $\vec{\xi}$ is the multiplet of chiral superfields,
$\vec{T}=\frac{1}{2}\vec{\sigma}$, and $\vec{\sigma}$ are the Pauli matrices.
The superfield multiplet obeys the constraint ${\rm det}(U)=f^2$ where $f^2$ is
the nonzero vacuum expectation value of the broken generator. And, an invariant
group element is then $V=U\bar{U}$ or $V^{-1}=\bar{U}^{-1} U^{-1}$. This
constitutes the $\lq\lq$standard" choice of coordinates $\xi$.

One can in principle evaluate all of the possible $SU(2)_L \times SU(2)_R$
terms eliminating the dependent invariants by use of integration by parts,
equations of motion, and general matrix relations. And, this is tenable for the
leading corrections in the action. They are
\bea
Tr(D^{\alpha}VD_{\alpha}V)\nonumber\\
Tr(D^{\alpha}V^{-1}D_{\alpha}V)\,.
\label{o3list}
\eea
Then the general action to leading order can be stated as
\bea
S(\vec{x})&=&\int d^4 x d^2 \theta^2 d^2 \bar{\theta}^2 \, a_m(Tr V)^m[1 + b_1
Tr(D^{\alpha}VD_{\alpha}V) \nonumber\\
& &+ b_2 Tr(D^{\alpha}V^{-1}D_{\alpha}V) + ... ]\,,
\eea
which restates previous results$^{[\ref{barnes}]}$ up to linear combinations
and total partial derivatives.

Because of the supersymmetric vectorial measure, dimensional analysis on the
component fields reveals that these invariants have dimension three. This means
that in this scheme the bosonic component fields $A$ carry dimension zero, the
fermionic fields $\psi$ carry dimension 1/2, and the auxiliary bosonic fields
$F$ carry dimension 1. This counting procedure is equivalent to assigning a
dimension of 1/2 to every supersymmetric derivative in the action while
assigning dimension 0 to each chiral superfield. So, unlike the bosonic case
where the possible dimensions can only be even, the supersymmetric case allows
for odd dimension terms. This implies that in the ordinary bosonic limit where
all fermionic fields are taken to be zero and the scalar fields are reduced to
Goldstone fields only, the odd dimension terms must vanish.

When this scheme is employed to find the subleading terms in this derivative
expansion, the number of possible terms becomes so large as to make the
calculation intractable. The problem arises not out of enumerating the terms
which is straightforward, but in establishing the independent set of terms.
This is in general a difficult question to answer.

Another formulation of the problem is to couch the model in geometrical terms.
If the purely non-supersymmetric bosonic model describes a global symmetry $G$
broken down to an invariant subgroup $H$, then a symmetric space can be formed
with the Goldstone fields corresponding to the broken generators acting as the
coordinates on the coset manifold $G/H$$^{[\ref{boulware}]}$. Then the usual
Killing vectors, metric, Riemann and Ricci tensors, etc. can be defined on this
manifold. Using this framework, Longhitano's two $SU(2)_L \times SU(2)_R$
invariant $O(p^4)$ terms can be restated as
\bea
g_{ij} g_{kl} \partial^{\mu} \phi^i \partial_{\mu} \phi^j \partial^{\nu} \phi^k
\partial_{\nu} \phi^l\nonumber\\
g_{ik} g_{jl} \partial^{\mu} \phi^i \partial_{\mu} \phi^j \partial^{\nu} \phi^k
\partial_{\nu} \phi^l
\label{longsterms}
\eea
where $g_{ij}$ is the metric of this manifold.

Extending this formulation to supersymmetry, it has been shown that the
manifold formed by the now complex Goldstone fields is in fact
K\"ahlerian$^{[\ref{helgason},\ref{zumino}]}$. This means that there exists a
real function $K(\phi,\bar{\phi})$ called the K\"ahler potential such that
\be
g_{i\bar{j}} = \frac{\partial^2 K(\phi,\bar{\phi})}{\partial \phi^i \partial
\bar{\phi^{\bar{j}}}}\,.
\ee
The form of the K\"ahler potential is restricted by the symmetry group $G$
whose action on the coordinates can be specified by Killing vectors where
$\delta_A \phi^i = A_A^i$. The choice of Killing vectors is not unique and in
fact there are an infinite set of vectors to choose from. However, different
choices of Killing vectors only correspond to different nonlinear realizations
of the global symmetry group which are physically equivalent. So, it is
convenient to choose the set of Killing vectors to be
\bea
A_A^i = \delta_A^a \epsilon_{aij} \phi^j + \delta_A^j(\delta_j^i +\phi^i
\phi^j) \nonumber\\
\bar{A}_A^{\bar{i}} = \delta_A^a \epsilon_{a\bar{i}\bar{j}}\bar{
\phi}^{\bar{j}} + \delta_A^{\bar{j}}(\delta_{\bar{j}}^{\bar{i}}
+\bar{\phi}^{\bar{i}} \bar{\phi}^{\bar{j}})
\eea
with the capital indices running over the full group $G$ while lower case
indices toward the beginning of the alphabet count over the invariant subgroup
$H$. Lower case indices toward the middle of the alphabet run over the
remaining broken generators or the broken coset $G/H$. This choice in turn
restricts the form of the K\"ahler potential to be$^{[\ref{buras}]}$
\be
K(\phi,\bar{\phi}) =
K'\left(\frac{1+\phi\cdot\bar{\phi}}{\sqrt{1+\phi^2}\sqrt{1+\bar{\phi}^2}}\right) + F(\phi^2) + \bar{F}(\bar{\phi}^2)\,.
\label{kahlereq}
\ee
Here, $F$ and $K'$ are arbitrary functions of their arguments with $K'$ obeying
the extra condition that $g_{i\bar{j}}(0,0) = \delta_{i\bar{j}}$. Note also
that this choice of Killing vectors does not correspond to the standard
coordinate choice of Longhitano's paper and will be denoted
$\lq\lq$geometrical" coordinates for lack of a better name. The connection
between the two coordinate systems will be elaborated on later in this paper.

To construct invariants then, the complete set of independent tensors must be
constructed and then contracted in every independent way. To this end, we make
an expansion in powers of momentum which, as discussed before, translates to an
expansion in number of covariant derivatives acting on the fields. For $O(p^3)$
terms, a basis set of tensors is formed from $D^{\alpha}\phi^i
D_{\alpha}\phi^j$ and its complex conjugate. This contracted with all possible
independent tensorial functions, $T_{ij}(\phi, \bar{\phi})$, results in the
complete set of invariants to this order. To be more specific, we also have to
consider the possibility of having a basis tensor of the form ${\cal
D}^2\phi^i$ so that the general tensor can be expressed as $T_i {\cal
D}^2\phi^i$. Here, ${\cal D}^2$ is the covariant derivative defined to be
${\cal D}^2 \phi^i = D^2 \phi^i + \omega^i_{jk} D^{\alpha}\phi^j D_{\alpha}
\phi^k$ where $\omega^i_{jk}$ is the Chiral connection defined in appendix A.
But sinc!
 e this falls under the supersymmetric vectorial measure, this invariant can be
transformed into $T_{i;j} D^{\alpha}\phi^i D_{\alpha}\phi^j$ via integration by
parts. In the notation used by this paper, $T_{i;j}$ denotes one covariant
derivative with respect to the field $\phi^j$ of the tensor $T_i$.

The tensorial objects $T_{ij}(\phi, \bar{\phi})$ can be formed from all
possible tensors, direct products, covariant coordinate derivatives, and
contractions among tensors. In a K\"ahler manifold with complex coordinates,
there are a number of candidates that must be considered including the Riemann
tensor $R^i_{j\bar{k}l}(\phi, \bar{\phi})$, K\"ahler metric $g_{i\bar{j}}(\phi,
\bar{\phi})$, the so called $\lq\lq$Chiral" metric $\gamma_{ij}(\phi)$,  the
$\lq\lq$Chiral" Riemann tensor $W_{ijkl}(\phi)$, and the Ricci tensor
$R_{ij}(\phi, \bar{\phi})$ \footnote{For a discussion of these tensors see
reference [\ref{love}].}. Another candidate that must be considered is chosen
from the arbitrary set of functions $K'(\phi, \bar{\phi})$ where the function
is required to be G-invariant. So we define the scalar function,
\be
K_0(\phi, \bar{\phi}) = \ln (1+\bar{\phi}\phi) - \frac{1}{2} \ln(1+\phi^2) -
\frac{1}{2} \ln(1+\bar{\phi}^2)
\ee
on which n covariant derivatives form an n-rank tensor. Notice that there are
an infinite number of choices of scalar quantities that can be defined for this
role since $K'(\phi,\bar{\phi})$ is an arbitrary function. However, covariant
derivatives of any of these choices can be related up to scalar factors to
$K_0(\phi, \bar{\phi})$. These scalar factors can then be absorbed in the
action at zero order. This procedure amounts to a field redefinition in the
action.

In addition, since every tensor transforming under this group can be expressed
in terms of $\delta_{ij}$,$\phi^i\phi^j$, and $\epsilon_{ijk}$, one would
expect a contribution to the set of tensors antisymmetric with respect to its
indices. Such objects can be constructed from the chiral and antichiral
vielbeins of the space which can be expressed as
\bea
e^{\,\underline{l}}_{\,m} =
\frac{1}{1+\phi^2}(\delta^{lm}+\epsilon_{lmn}\phi^n) \nonumber\\
\bar{e}^{\,\underline{l}}_{\,\bar{m}} =
\frac{1}{1+\bar{\phi}^2}(\delta^{lm}+\epsilon_{lmn}\bar{\phi}^n)\,.
\eea
The underlined indices refer to a tangent space index which transforms linearly
while the remaining index transforms nonlinearly under the broken symmetries.
Given this, we define the tensor
\be
T_{ijk}=\epsilon_{mnp} e^{\underline{m}}_{\,\, i} e^{\underline{n}}_{\,\, j}
e^{\underline{p}}_{\,\, k}
\ee
along with its complex conjugate which has a component that is proportional to
$\epsilon_{ijk}$ and so totally antisymmetric under interchange of i,j, or k.

{}From this list of tensors, it can be proven that the independent set of
tensors up to direct products are
\bea
&K_{0;i}&\nonumber\\
\gamma_{ij}&\equiv&A_{Ai} A_{Aj} \nonumber\\
g_{i\bar{j}}&\equiv&K_{0;i\bar{j}} \nonumber\\
T_{ijk}&\equiv&\epsilon_{mnp} e^{\underline{m}}_{\,\, i}
e^{\underline{n}}_{\,\, j} e^{\underline{p}}_{\,\, k} \nonumber\\
S_{ij}&\equiv&T_{ijk} K_0^{\, ;k} \nonumber\\
A_{ij\bar{k}}&\equiv&T_{ijm} \gamma ^{mn} g_{\,n\bar{k}} \nonumber\\
I_{j\bar{k}}&\equiv&A_{ij\bar{k}} K_0^{\, ;i}
\label{ntensor}
\eea
along with complex conjugates where necessary \footnote{For a more complete
description of these tensors, see appendix A.}. No further contractions or
covariant derivatives can be taken that generate an independent tensor.
Therefore, the possible tensors at any order can be easily specified as either
one of the these tensors or a direct product of a combination of them. As a
result,  the $O(p^3)$ terms can be immediately written down as
\bea
K_{0;i} K_{0;j} D^{\alpha}\phi^i D_{\alpha}\phi^j \nonumber\\
\gamma_{ij} D^{\alpha}\phi^i D_{\alpha}\phi^j
\eea
along with their complex conjugates. Note that it is assumed that each
invariant is associated with a complex coefficient so the action will remain
hermitian. This list reproduces the result of equation (\ref{o3list}) in
geometric form.

It is worth noting at this point the connection between the geometrical and the
matrix (standard coordinate) notations. These two cases really represent just a
change in non-linear realization of the global symmetry group. Thus, given a
suitable redefinition of the Goldstone fields $\phi^i$, a translation can be
made$^{[\ref{buras}]}$. In this case if we define the scalar invariant
\be
a(\phi,\bar{\phi}) =
\frac{1+\bar{\phi}\phi}{\sqrt{1+\phi^2}\sqrt{1+\bar{\phi}^2}}\,,
\ee
then
\be
Tr(V) = 2 a(\phi,\bar{\phi})\,.
\ee
Furthermore, defining $\xi^i$ to be a field in standard coordinates and
$\phi^i$ to be a field in the geometrical notation of this paper, translations
of any tensorial quantity can be made with the assignment $\xi^i=\frac{2}{\phi}
\arctan(\phi) \phi^i$. Note that $\phi = \sqrt{\phi \cdot \phi}$.

Now, we can proceed to the subleading $O(p^4)$ terms. This case adds a few
complications in that there are four covariant derivatives acting on the fields
as well as the four derivatives from the vector measure which allows for more
base tensors as well as more complicated tensors $T_{ijkl}(\phi,\bar{\phi})$.
However, the tensors $T_{ijkl}(\phi,\bar{\phi})$ are still formed by direct
products of the original independent tensors of equation (\ref{ntensor}). In
addition, it must be realized that the derivatives acting on the fields are
covariant derivatives which require extra care. For example, the Grassmannian
property $D^3\phi = 0$ does not hold for covariant derivatives ${\cal D}^3\phi
\neq 0$. Another consideration is the fact that the bases at this order are not
independent due to integration by parts. Nevertheless, the evaluation of terms
are still tractable in the geometrical notation and in fact there are 42 terms:
\be
\begin{array}{ll}
1.\, K_{0;i}K_{0;j}{\cal D}^2\phi^i {\cal D}^2 \phi^j &
14.\, K_{0;i} I_{j\bar{k}} \bar{D}^{\dot{\alpha}} D^{\alpha}\phi^i
D_{\alpha}\phi^j \bar{D}_{\dot{\alpha}}\bar{\phi}^{\bar{k}} \\
2.\, \gamma_{ij}{\cal D}^2\phi^i {\cal D}^2 \phi^j &
15.\, K_{0;i}g_{j\bar{k}} \bar{D}^{\dot{\alpha}} D^{\alpha}\phi^i
D_{\alpha}\phi^j \bar{D}_{\dot{\alpha}}\bar{\phi}^{\bar{k}} \\
3.\, K_{0;i}K_{0;\bar{j}}{\cal D}^2\phi^i \bar{{\cal D}}^2 \bar{\phi}^{\bar{j}}
&
16.\, K_{0;i}K_{0;j}K_{0;\bar{k}}K_{0;\bar{l}} D^{\alpha}\phi^i
D_{\alpha}\phi^j \bar{D}_{\dot{\alpha}}\bar{\phi}^{\bar{k}}
\bar{D}^{\dot{\alpha}}\bar{\phi}^{\bar{l}} \\
4.\, g_{i\bar{j}}{\cal D}^2\phi^i \bar{{\cal D}}^2 \bar{\phi}^{\bar{j}} &
17.\, g_{i\bar{k}}g_{j\bar{l}} D^{\alpha}\phi^i D_{\alpha}\phi^j
\bar{D}_{\dot{\alpha}}\bar{\phi}^{\bar{k}}
\bar{D}^{\dot{\alpha}}\bar{\phi}^{\bar{l}} \\
5.\, I_{i\bar{j}} {\cal D}^2\phi^i \bar{{\cal D}}^2 \bar{\phi}^{\bar{j}} &
18.\, K_{0;i}K_{0;j}\bar{\gamma}_{\bar{k}\bar{l}} D^{\alpha}\phi^i
D_{\alpha}\phi^j \bar{D}_{\dot{\alpha}}\bar{\phi}^{\bar{k}}
\bar{D}^{\dot{\alpha}}\bar{\phi}^{\bar{l}} \\
6.\, K_{0;i}\gamma_{jk} {\cal D}^2\phi^i D^{\alpha}\phi^j D_{\alpha}\phi^k &
19.\, K_{0;i}K_{0;\bar{k}}g_{j\bar{l}} D^{\alpha}\phi^i D_{\alpha}\phi^j
\bar{D}_{\dot{\alpha}}\bar{\phi}^{\bar{k}}
\bar{D}^{\dot{\alpha}}\bar{\phi}^{\bar{l}} \\
7.\, K_{0;i}K_{0;j}K_{0;k} {\cal D}^2\phi^i D^{\alpha}\phi^j D_{\alpha}\phi^k &
20.\, \gamma_{ij}\bar{\gamma}_{\bar{k}\bar{l}} D^{\alpha}\phi^i
D_{\alpha}\phi^j \bar{D}_{\dot{\alpha}}\bar{\phi}^{\bar{k}}
\bar{D}^{\dot{\alpha}}\bar{\phi}^{\bar{l}} \\
8. \,\gamma_{ij} K_{0;\bar{k}} D^{\alpha}\phi^i D_{\alpha}\phi^j\bar{{\cal
D}}^2 \bar{\phi}^{\bar{k}} &
21.\, K_{0;i}K_{0;\bar{k}}I_{j\bar{l}} D^{\alpha}\phi^i D_{\alpha}\phi^j
\bar{D}_{\dot{\alpha}}\bar{\phi}^{\bar{k}}
\bar{D}^{\dot{\alpha}}\bar{\phi}^{\bar{l}} \\
9.\, K_{0;i}g_{j\bar{k}} D^{\alpha}\phi^i D_{\alpha}\phi^j\bar{{\cal D}}^2
\bar{\phi}^{\bar{k}} &
22.\, I_{i\bar{k}}I_{j\bar{l}} D^{\alpha}\phi^i D_{\alpha}\phi^j
\bar{D}_{\dot{\alpha}}\bar{\phi}^{\bar{k}}
\bar{D}^{\dot{\alpha}}\bar{\phi}^{\bar{l}} \\
10.\, K_{0;i}K_{0;j}K_{0;\bar{k}} D^{\alpha}\phi^i D_{\alpha}\phi^j\bar{{\cal
D}}^2 \bar{\phi}^{\bar{k}} &
23.\, I_{i\bar{k}} g_{j\bar{l}} D^{\alpha}\phi^i D_{\alpha}\phi^j
\bar{D}_{\dot{\alpha}}\bar{\phi}^{\bar{k}}
\bar{D}^{\dot{\alpha}}\bar{\phi}^{\bar{l}} \\
11.\, K_{0;i} I_{j\bar{k}} D^{\alpha}\phi^i D_{\alpha}\phi^j\bar{{\cal D}}^2
\bar{\phi}^{\bar{k}} &
24.\, K_{0;i}K_{0;j}\gamma_{kl} D^{\alpha}\phi^i D_{\alpha}\phi^j
D^{\beta}\phi^k D_{\beta}\phi^l \\
12.\, A_{ij\bar{k}} \bar{D}^{\dot{\alpha}} D^{\alpha}\phi^i D_{\alpha}\phi^j
\bar{D}_{\dot{\alpha}}\bar{\phi}^{\bar{k}} &
25.\, \gamma_{ij}\gamma_{kl} D^{\alpha}\phi^i D_{\alpha}\phi^j D^{\beta}\phi^k
D_{\beta}\phi^l \\
13.\, S_{ij}K_{0;\bar{k}} \bar{D}^{\dot{\alpha}} D^{\alpha}\phi^i
D_{\alpha}\phi^j \bar{D}_{\dot{\alpha}}\bar{\phi}^{\bar{k}} &
26.\, S_{ik}S_{jl} D^{\alpha}\phi^i D_{\alpha}\phi^j D^{\beta}\phi^k
D_{\beta}\phi^l
\end{array}
\label{invlist}
\ee
The other terms come from complex conjugates where necessary. As a check, we
require this list to recover Longhitano's result in the bosonic limit and it
can be shown that we do indeed recover the invariants of equation
(\ref{longsterms}) when this constraint is implemented. Also in this limit, the
$O(p^3)$ terms vanish as expected.

It should be noted that nothing has been said about the use of field equations
to reduce the number of independent operators of the problem on shell. This has
been done because no assumption has been made as to the order in momentum to
which this expansion is to be truncated which in turn establishes an order to
which the field equations are valid. It should be stressed that terms
eliminated with the use of a lower order field equation will not necessarily be
eliminated in a higher order calculation. With this in mind, we choose the
expansion to truncate at $O(p^4)$ which establishes the field equations as
\bea
{\cal D}^2 \phi^i &=& 2 K_{0;j} D^{\alpha}\phi^i D_{\alpha}\phi^j + O(p^4)
\nonumber\\
{\bar{\cal D}}^2 \bar{\phi}^{\bar{i}} &=& 2 K_{0;\bar{j}}
\bar{D}_{\dot{\alpha}}\bar{\phi}^{\bar{i}}
\bar{D}^{\dot{\alpha}}\bar{\phi}^{\bar{j}} +O(p^4)\,.
\eea
This condition acts to eliminate all of the $O(p^4)$ terms containing second
derivatives. Furthermore, it and its complex conjugate establish a relationship
between the $O(p^3)$ and $O(p^4)$ terms that removes two more operators on
shell. Thus, we choose to eliminate the terms
\bea
K_{0;i}K_{0;j}\bar{\gamma}_{\bar{k}\bar{l}} D^{\alpha}\phi^i D_{\alpha}\phi^j
\bar{D}_{\dot{\alpha}}\bar{\phi}^{\bar{k}}
\bar{D}^{\dot{\alpha}}\bar{\phi}^{\bar{l}} \nonumber \\
\gamma_{ij}K_{0;\bar{k}}K_{0;\bar{l}} D^{\alpha}\phi^i D_{\alpha}\phi^j
\bar{D}_{\dot{\alpha}}\bar{\phi}^{\bar{k}}
\bar{D}^{\dot{\alpha}}\bar{\phi}^{\bar{l}}\,.
\eea
It must be noted as well that these equations constitute a constraint on the
superfields only under the vectorial measure. Since the remaining terms in the
action still fall under the vectorial measure, there are components that can
still be eliminated with the field equation. However since this would destroy
the explicit supersymmetric construction of the Lagrangian, we refrain from
applying this condition here. So the field equation further reduces the list to
a final result of 21 invariants.
\be
\begin{array}{l}
1.\quad K_{0;i}g_{j\bar{k}} \bar{D}^{\dot{\alpha}} D^{\alpha}\phi^i
D_{\alpha}\phi^j \bar{D}_{\dot{\alpha}}\bar{\phi}^{\bar{k}} \\
2.\quad A_{ij\bar{k}} \bar{D}^{\dot{\alpha}} D^{\alpha}\phi^i D_{\alpha}\phi^j
\bar{D}_{\dot{\alpha}}\bar{\phi}^{\bar{k}} \\
3.\quad S_{ij}K_{0;\bar{k}} \bar{D}^{\dot{\alpha}} D^{\alpha}\phi^i
D_{\alpha}\phi^j \bar{D}_{\dot{\alpha}}\bar{\phi}^{\bar{k}} \\
4.\quad K_{0;i} I_{j\bar{k}} \bar{D}^{\dot{\alpha}} D^{\alpha}\phi^i
D_{\alpha}\phi^j \bar{D}_{\dot{\alpha}}\bar{\phi}^{\bar{k}} \\
5.\quad K_{0;\bar{i}}g_{\bar{j}k}
D^{\alpha}\bar{D}^{\dot{\alpha}}\bar{\phi}^{\bar{i}}
\bar{D}_{\dot{\alpha}}\bar{\phi}^{\bar{j}} D_{\alpha}\phi^k \\
6.\quad \bar{A}_{\bar{i}\bar{j}k}
D^{\alpha}\bar{D}^{\dot{\alpha}}\bar{\phi}^{\bar{i}}
\bar{D}_{\dot{\alpha}}\bar{\phi}^{\bar{j}} D_{\alpha}\phi^k \\
7.\quad \bar{S}_{\bar{i}\bar{j}}K_{0;k}
D^{\alpha}\bar{D}^{\dot{\alpha}}\bar{\phi}^{\bar{i}}
\bar{D}_{\dot{\alpha}}\bar{\phi}^{\bar{j}} D_{\alpha}\phi^k \\
8.\quad K_{0;\bar{i}} I_{\bar{j}k}
D^{\alpha}\bar{D}^{\dot{\alpha}}\bar{\phi}^{\bar{i}}
\bar{D}_{\dot{\alpha}}\bar{\phi}^{\bar{j}} D_{\alpha}\phi^k \\
9.\quad K_{0;i}K_{0;j}K_{0;\bar{k}}K_{0;\bar{l}} D^{\alpha}\phi^i
D_{\alpha}\phi^j \bar{D}_{\dot{\alpha}}\bar{\phi}^{\bar{k}}
\bar{D}^{\dot{\alpha}}\bar{\phi}^{\bar{l}} \\
10.\quad g_{i\bar{k}}g_{j\bar{l}} D^{\alpha}\phi^i D_{\alpha}\phi^j
\bar{D}_{\dot{\alpha}}\bar{\phi}^{\bar{k}}
\bar{D}^{\dot{\alpha}}\bar{\phi}^{\bar{l}} \\
11.\quad K_{0;i}K_{0;\bar{k}}g_{j\bar{l}} D^{\alpha}\phi^i D_{\alpha}\phi^j
\bar{D}_{\dot{\alpha}}\bar{\phi}^{\bar{k}}
\bar{D}^{\dot{\alpha}}\bar{\phi}^{\bar{l}} \\
12.\quad \gamma_{ij}\bar{\gamma}_{\bar{k}\bar{l}} D^{\alpha}\phi^i
D_{\alpha}\phi^j \bar{D}_{\dot{\alpha}}\bar{\phi}^{\bar{k}}
\bar{D}^{\dot{\alpha}}\bar{\phi}^{\bar{l}}  \\
13.\quad K_{0;i}K_{0;\bar{k}}I_{j\bar{l}} D^{\alpha}\phi^i D_{\alpha}\phi^j
\bar{D}_{\dot{\alpha}}\bar{\phi}^{\bar{k}}
\bar{D}^{\dot{\alpha}}\bar{\phi}^{\bar{l}} \\
14.\quad I_{i\bar{k}}I_{j\bar{l}} D^{\alpha}\phi^i D_{\alpha}\phi^j
\bar{D}_{\dot{\alpha}}\bar{\phi}^{\bar{k}}
\bar{D}^{\dot{\alpha}}\bar{\phi}^{\bar{l}} \\
15.\quad I_{i\bar{k}} g_{j\bar{l}} D^{\alpha}\phi^i D_{\alpha}\phi^j
\bar{D}_{\dot{\alpha}}\bar{\phi}^{\bar{k}}
\bar{D}^{\dot{\alpha}}\bar{\phi}^{\bar{l}} \\
16.\quad K_{0;i}K_{0;j}\gamma_{kl} D^{\alpha}\phi^i D_{\alpha}\phi^j
D^{\beta}\phi^k D_{\beta}\phi^l \\
17.\quad \gamma_{ij}\gamma_{kl} D^{\alpha}\phi^i D_{\alpha}\phi^j
D^{\beta}\phi^k D_{\beta}\phi^l \\
18.\quad S_{ik}S_{jl} D^{\alpha}\phi^i D_{\alpha}\phi^j D^{\beta}\phi^k
D_{\beta}\phi^l \\
19.\quad K_{0;\bar{i}}K_{0;\bar{j}}\bar{\gamma}_{\bar{k}\bar{l}}
\bar{D}_{\dot{\alpha}}\bar{\phi}^{\bar{i}}
\bar{D}^{\dot{\alpha}}\bar{\phi}^{\bar{j}}
\bar{D}_{\dot{\beta}}\bar{\phi}^{\bar{k}}
\bar{D}^{\dot{\beta}}\bar{\phi}^{\bar{l}} \\
20.\quad \bar{\gamma}_{\bar{i}\bar{j}}\bar{\gamma}_{\bar{k}\bar{l}}
\bar{D}_{\dot{\alpha}}\bar{\phi}^{\bar{i}}
\bar{D}^{\dot{\alpha}}\bar{\phi}^{\bar{j}}
\bar{D}_{\dot{\beta}}\bar{\phi}^{\bar{k}}
\bar{D}^{\dot{\beta}}\bar{\phi}^{\bar{l}} \\
21.\quad \bar{S}_{\bar{i}\bar{k}}\bar{S}_{\bar{j}\bar{l}}
\bar{D}_{\dot{\alpha}}\bar{\phi}^{\bar{i}}
\bar{D}^{\dot{\alpha}}\bar{\phi}^{\bar{j}}
\bar{D}_{\dot{\beta}}\bar{\phi}^{\bar{k}}
\bar{D}^{\dot{\beta}}\bar{\phi}^{\bar{l}}
\end{array}
\ee
This list is considerably longer than the 10 $O(p^4)$ terms of reference
[\ref{barnes}]\footnote{See appendix B for a discussion of this.}. It
represents a comprehensive list of the number of independent operators to this
order which are unexpectedly large in comparison with the Electroweak case.

As a next step, one could constrain the model to obey a $SU(2)_L \times U(1)$
symmetry retaining terms to the same $O(p^4)$ order. Such a model would be an
extension of the model presented by reference [\ref{clark}] and lies within
current experimental constraints on the extensions of the Standard Model. Since
the invariants listed in this paper also obey this new symmetry, one would
expect a more complex Lagrangian as a result. This exercise will be left for a
future paper.

\newpage
\centerline{\large\bf Appendix A:}
\centerline{\large\bf Tensor construction in the geometrical basis}

\setcounter{equation}{0}
\renewcommand{\theequation}{A.\arabic{equation}}

The key to successfully constructing a set of invariants relies on constructing
the most general set of tensors such that only direct products give rise to new
tensors. In this way, much of the work of determining independence of possible
candidates can be partitioned into a question of tensorial independence and
connections among invariants by integration by parts. The easiest way to
approach the problem of tensorial independence is to work in the simplest
coordinate system one can find so that the dependences are not overly
complicated. The geometrical coordinates used in this paper result in the
simplest forms of the tensors and were chosen precisely for that reason.

We start with the simplest tensorial quantity that can be defined for the case
of $SU(2)_L \times SU(2)_R$ invariance which is the scalar. As seen by equation
(\ref{kahlereq}), there is only one combination of the fields $\phi$ and
$\bar{\phi}$ that results in a scalar invariant aside from arbitrary powers of
the invariant. So we define the scalar
\be
a \equiv \frac{\sqrt{\beta \bar{\beta}}}{\alpha}
\ee
where
\bea
\beta &=& \frac{1}{1+\phi^2} \nonumber\\
\bar{\beta} &=& \frac{1}{1+\bar{\phi}^2} \nonumber\\
\alpha &=& \frac{1}{1+\phi\cdot \bar{\phi}} \,.
\eea

So the function $K_0$ can be simply stated as $K_0 = \ln(a)$. Thus, one
derivative (or one covariant derivative) creates a rank 1 tensor:
\be
K_{0;i} = \alpha \bar{\phi}^i - \beta \phi^i\,.
\ee
Applying another derivative with respect to the complex conjugated field
results in the metric of the space:
\be
g_{\bar{i}j} \equiv K_{0;\bar{i}j} = \alpha (\delta^{\bar{i}j} - \alpha
\phi^{\bar{i}} \bar{\phi^j})\,.
\ee
Notice that this metric is not formed from the contraction of two Killing
vectors. Thus, we must define another two index tensor which we shall denote
the Chiral metric:
\be
\gamma_{ij} \equiv A_{Ai} A_{Aj} = \beta(\delta^{ij} - \beta \phi^i \phi^j) \,.
\ee

Taking covariant derivatives of a tensorial quantity creates a higher rank
tensor. Unfortunately, the covariant derivative is not uniquely defined beyond
one derivative and in fact there are two ways to construct a covariant
derivative. The first employs the usual connection defined for the space which
is defined from the K\"ahler potential.
\be
\Gamma^i_{jk} \equiv g^{i\bar{n}}K_{0,\bar{n}jk} = -\alpha(\bar{\phi}^k
\delta^i_j + \bar{\phi}^j \delta^i_k)\,.
\ee
But another connection can be defined which shall be called the Chiral
connection$^{[\ref{love}]}$. It is defined by
\be
\omega^i_{jk} \equiv \frac{1}{2} \gamma^{in}(\gamma_{nk,j} + \gamma_{nj,k} -
\gamma_{jk,n}) = -\beta(\phi^k \delta^i_j + \phi^j \delta^i_k)\,.
\ee
In these equations, a comma denotes an ordinary derivative with respect to the
field. Both of these objects deserve the apellation connection since they
transform as Christoffel symbols. Thus, two covariant derivatives can be
defined. For example, a covariant derivative of a first rank tensor could be
\be
T_{i;j} \equiv T_{i,j} - \omega^m_{ij}T_m
\ee
or
\be
T_{i;j'} \equiv T_{i,j} - \Gamma^m_{ij}T_m \,.
\ee
However, it can be shown that $T_{i;j'} = T_{i;j} + K_{0;i}T_j + K_{0,j}T_i$
and so the two definitions are related. This procedure can be implemented to
all orders of covariant derivatives which allows us the freedom to choose one
definition. In this paper, we chose to use covariant differentiation defined by
the Chiral connection.

If we apply covariant derivatives to the given tensors defined so far, we find
that this set of tensors closes up to direct products. For example, we have
\bea
\gamma_{ij;k}&=&0 \nonumber \\
K_{0;ij} &=& -K_{0;i}K_{0,j}-\gamma_{ij} \nonumber \\
g_{\bar{i}j;k}&=&-g_{\bar{i}j}K_{0;k}-g_{\bar{i}k}K_{0;j}\,.
\eea
We have not addressed the issue of contractions among these tensors. However,
it is seen that these tensors do in fact contract up in such a way as to
produce another object in the list or direct products thereof. This indicates
we have a $\lq\lq$basis" set of tensors that are independent and can be used as
building blocks of more complex tensors via direct products.

However, this is not all of the possible tensor expressions. If one considers a
general function that transforms as a tensor under the group $SU(2)_L \times
SU(2)_R$, one finds that it can be expressed as combinations of
$\phi^i$,$\delta^{ij}$, and $\epsilon_{ijk}$. This arises from the fact that
traces of Pauli matrices lead to functions of this sort. So one might expect a
tensorial quantity arising from a $\epsilon_{ijk}$ component. And indeed, such
tensors have been expressed in the main body of this paper all of which arise
from the tensor $T_{ijk}$. There, we defined it in terms of the vielbeins of
the space, however for the purposes of the appendices, it would be simpler to
extract out the essential quantity of $T_{ijk}$; the component proportional to
$\epsilon_{ijk}$. In this spirit, we redefine $T_{ijk}$ to be
\be
T_{ijk} = \beta^2 \epsilon_{ijk}\,.
\ee
The other antisymmetric tensors arise naturally from contractions of $T_{ijk}$
with our previously defined set of tensors and they have the form:
\bea
S_{ij}&=&\alpha\beta\epsilon_{ijk}(\bar{\phi}^k - \phi^k) \nonumber \\
A_{ij\bar{k}} &=& \frac{1}{a^2} \beta^2 \bar{\beta} (\epsilon_{ij\bar{k}} +
\epsilon_{ijm}\phi^k\phi^m + \epsilon_{jkm}\phi^i\bar{\phi}^m -
\epsilon_{ikm}\phi^j\bar{\phi}^m) \nonumber \\
I_{i\bar{j}} &=& \beta \bar{\beta} \epsilon_{i\bar{j}k} (\bar{\phi}^k - \phi^k)
\eea

As a whole, these seven tensors constitute a closed set; closed in the sense
that no contractions or covariant derivatives result in a new tensor.
Therefore, these seven tensors constitute the necessary building blocks to form
invariants under this group.

As a final confirmation of the completeness of this set, we can evaluate some
basic tensors of the manifold. The Riemann tensor can be constructed
as$^{[\ref{shore2}]}$
\be
R_{\bar{i}j\bar{k}l} \equiv g_{\bar{i}n}\Gamma^n_{jl,\bar{k}} = -[g_{\bar{i}j}
g_{\bar{k}l}+g_{\bar{i}l}g_{\bar{k}j}] \,.
\ee
Similarly, the Chiral Riemann tensor has the form
\bea
W_{ijkl} &\equiv& \gamma_{im}[\omega^m_{jl,k} - \omega^m_{jk,l} +
\omega^m_{nk}\omega^n_{jl} - \omega^m_{nl}\omega^n_{jk}] \nonumber \\
&=& \gamma_{ik}\gamma_{jl} - \gamma_{il}\gamma_{jk}\,.
\eea
And finally, the Ricci tensor for a general K\"ahler manifold can be derived
from the metric of the space$^{[\ref{alvarez}]}$:
\bea
R_{i\bar{j}} &\equiv& \frac{\partial^2}{\partial \phi^i \partial
\bar{\phi}^{\bar{j}}} \ln {\rm det}(g_{\bar{k}l}) \nonumber \\
&=& -4 g_{i\bar{j}}\,.
\eea

\newpage
\centerline{\large\bf Appendix B:}
\centerline{\large\bf A translation of coordinates}

\setcounter{equation}{0}
\renewcommand{\theequation}{B.\arabic{equation}}

Although the standard coordinates are typically written in terms of the Pauli
matrices, this by no means disallows a reformulation of the coordinates into a
geometrical way of looking at them. The unitary matrix $exp(i \vec{T}\cdot
\vec{\xi})$ can be written as $U= cos(\frac{\xi}{2}) + \frac{2i}{\xi}
sin(\frac{\xi}{2}) \vec{\xi}\cdot\vec{T}$. So if we were to proceed as in
Appendix A, we would take
\be
Tr(U\bar{U})=2\left[
cos(\frac{\xi}{2})cos(\frac{\bar{\xi}}{2})+sin(\frac{\xi}{2})
sin(\frac{\bar{\xi}}{2}) \frac{\bar{\xi}\cdot\xi}{\bar{\xi}\xi}\right]
\ee
and define that as our scalar invariant, take covariant derivatives, and
proceed to build a tensor list as before. This gives rise to a completely
equivalent $\lq\lq$geometrical" interpretation of the action. For example, the
Chiral metric defined in appendix A has the form in standard coordinates:
\bea
\gamma_{ij}(\xi) &=& \frac{\partial \xi^i}{\partial \phi^m} \frac{\partial
\xi^j}{\partial \phi^n} \gamma^{mn}(\phi) \nonumber \\
&=& \delta^{ij} \frac{t^2}{(1+t^2)\xi^2} -
\xi^i\xi^j\left(\frac{t^2}{(1+t^2)\xi^4} - \frac{1}{4\xi^2}\right)
\eea
where $t=\tan(\frac{\xi}{2})$. In general, if we parametrized the Killing
vectors of a field $\phi$ to be
\be
A_A^i = \delta_A^a \epsilon_{aij} \phi^j + \delta_A^j\left[\delta_j^i g(\phi)
+\phi^i \phi^j h(\phi)\right]
\ee
then a general Chiral metric can be defined as
\be
\gamma^{ij} = \delta^{ij}(g^2 + \phi^2) + \phi^i \phi^j (h^2 \phi^2 + 2 gh
-1)\,.
\ee
An examination of this equation reveals that the choice of $f=1$ and $g=1$
gives the metric in its simplest form which is also the geometrical choice of
coordinates used in this paper.

We can use this kind of redefinition to formulate a translation from the
standard coordinates to the geometric coordinates. We have already seen that
$Tr(V)=2a$ is a relation between the two field definitions accomplished via the
field redefinition $\xi^i=\frac{2}{\phi} \arctan(\phi) \phi^i$. So, using this
translation directly on $U$ and $\bar{U}$ leads to the result:
\bea
U&=&\sqrt{\beta}(1+2i\vec{\phi}\cdot \vec{T}) \nonumber \\
\bar{U}&=&\sqrt{\bar{\beta}}(1-2i \vec{\bar{\phi}}\cdot \vec{T})\,.
\eea
So in principle, one can take any standard coordinate invariant and translate
it into the geometrical coordinate equivalent by taking the appropriate
derivatives and traces. One could, for example, take the list generated by
reference [\ref{barnes}] and find the geometrical coordinate equivalent. And
indeed, a couple of the simplest terms are:
\bea
\bar{D}^2 Tr\bar{Z}\,Tr\bar{Z}\,Tr\bar{Z} &=& -\frac{1}{8} \bar{D}^2 D^2
a^3(K_{0;i}{\cal D}^2 \phi^i - \gamma_{ij} D^{\alpha}\phi^i D_{\alpha}\phi^j)
\nonumber \\
& & \quad\times (K_{0;k}{\cal D}^2 \phi^k - \gamma_{kl} D^{\alpha}\phi^k
D_{\alpha}\phi^l) \nonumber \\
\bar{D}^2 D^2 TrZ\,Tr\bar{Z} &=& \frac{1}{4}\bar{D}^2 D^2 a^2 (K_{0;i}{\cal
D}^2 \phi^i - \gamma_{ij} D^{\alpha}\phi^i D_{\alpha}\phi^j) \nonumber \\
& & \quad\times (K_{0;\bar{k}}{\cal D}^2 \phi^{\bar{k}} -
\bar{\gamma}_{\bar{k}\bar{l}} \bar{D}_{\dot{\alpha}}\phi^{\bar{k}}
\bar{D}^{\dot{\alpha}}\phi^{\bar{l}})
\eea
where
\bea
Z&=& -\frac{1}{4} \bar{D}^2 V \nonumber \\
W_{\alpha}&=& -\frac{1}{4}\bar{D}^2 D_{\alpha} V\,.
\eea
As one can see, the two coordinates heavily mix upon translation making a one
to one correspondence difficult. However, one can count the invariants
generated by the list, and with a little effort, one finds that the 10 $O(p^4)$
terms\footnote{Actually only 8 of the 10 terms are independent since two of
them can be eliminated with the use of ${\rm det} M = \frac{1}{2}[Tr(M) Tr(M) -
Tr(M^2)]$ where $M$ is a general $2 \times 2$ matrix.} involve 23 of the
invariants listed in equation (\ref{invlist}).

One might be interested to see if any of the other invariants not part of this
list can be generated in the standard coordinates and upon consideration it can
be shown that this is the case. There are various combinations of matrices that
can be traced to immediately see a tensorial component. A short and incomplete
list is:
\bea
Tr(D^{\alpha}V) &=& 2a K_{0;i} D^{\alpha} \phi^i \nonumber \\
Tr(D^{\alpha}V D_{\alpha}V) &=& \left[4a^2 K_{0;i} K_{0;j} - 2 \gamma_{ij}
\right] D^{\alpha} \phi^i D_{\alpha} \phi^j \nonumber \\
Tr(VD^{\alpha}V D^{\beta}V) &=& -2a S_{ij} \,D^{\alpha} \phi^i D^{\beta} \phi^j
+ ...\nonumber \\
Tr(\bar{D}^{\dot{\alpha}} D^{\alpha} V) &=& 2a\left[ K_{0;i} K_{0;\bar{j}} +
g_{i\bar{j}} \right] \bar{D}^{\dot{\alpha}} \bar{\phi}^{\bar{j}} D^{\alpha}
\phi^i + 2aK_{0;i} \bar{D}^{\dot{\alpha}}D^{\alpha} \phi^i \nonumber \\
Tr(V\bar{D}^{\dot{\alpha}} D^{\alpha} V) &=& 2 I_{i\bar{j}}\,
\bar{D}^{\dot{\alpha}} \bar{\phi}^{\bar{j}} D^{\alpha} \phi^i + ... \,.
\eea
{}From this, it becomes evident that many of the invariants listed in equation
(\ref{invlist}) can be immediately generated including, for example, a couple
that are not created from combinations of the matrices $Z$ and $W_{\alpha}$:
\bea
Tr(VD^{\alpha}V D^{\beta}V)\,Tr(VD_{\alpha}V D_{\beta}V) &=& 4a S_{ik}S_{jl}
D^{\alpha} \phi^i D_{\alpha} \phi^j D^{\beta} \phi^k D_{\beta} \phi^l
+...\nonumber \\
Tr(V\bar{D}^{\dot{\alpha}} D^{\alpha} V)\,Tr(\bar{D}_{\dot{\alpha}} D_{\alpha}
V) &=& -4a K_{0;i}I_{j\bar{k}}\,\bar{D}^{\dot{\alpha}}D^{\alpha} \phi^i
D_{\alpha} \phi^j \bar{D}_{\dot{\alpha}} \bar{\phi}^{\bar{k}} +... \,.\nonumber
\\
\eea
So it can be concluded that all independent invariants of this group cannot be
written solely as functions of  $Z$ and $W_{\alpha}$.

\bigskip

\newpage
\noindent
{\bf References}
\begin{enumerate}

\item \label{long}
A.C. Longhitano,Nucl. Phys. {\bf B188}(1980)118.

\item \label{appel}
T. Appelquist and Guo-Hong Wu, Phys. Rev. {\bf D48}(1993)48.

\item \label{barnes}
K.J.Barnes,D.A Ross and R. D. Simmons,Phys. Lett. {\bf B338}(1994)457.

\item \label{boulware}
D.G. Boulware and L.S. Brown, Annals of Physics {\bf138}(1982)393.

\item \label{helgason}
S. Helgason,Differential geometry, Lie groups and symmetric spaces (Academic
Press, New York,1978).

\item \label{zumino}
B. Zumino,Phys. Lett. {\bf 87B}(1979)203.

\item \label{shore}
G.M. Shore, Nucl. Phys. {\bf B320}(1989)202.

\item \label{buras}
A.J. Buras and W. Slominski,Nucl. Phys. {\bf B223}(1983)157.

\item \label{love}
T.E. Clark and S.T. Love, Nucl. Phys. {\bf B301}(1988)439.

\item \label{clark}
T.E. Clark and W.T.A. ter Veldhuis, Nucl. Phys. {\bf B426}(1994)385.

\item \label{shore2}
G.M. Shore, Nucl. Phys {\bf B334}(1990)172.

\item \label{alvarez}
Luis Alvarez-Gaum\'e and Daniel Z. Freedman, Phys. Lett. {\bf 94B}(1980)171.

\end{enumerate}
\end{document}